\documentclass[conference,9pt]{IEEEtran}

\usepackage[2023, cc-by, pages={3}{4}, url=https://osda.ws/r/UY4P1]{osda}

\IEEEoverridecommandlockouts
\usepackage{cite}
\usepackage{hyperref} 
\usepackage[noend]{algpseudocode}
\usepackage{caption}
\usepackage{graphicx}
\usepackage{xcolor}
\usepackage{booktabs}
\usepackage{listings}

\def\BibTeX{{\rm B\kern-.05em{\sc i\kern-.025em b}\kern-.08em
    T\kern-.1667em\lower.7ex\hbox{E}\kern-.125emX}}
\begin{document}

%%%%%%%%%%%%%%%%%%%%%%%%%%%%%%%%%%%%
\newcommand{\refsec}[1]{Section~\ref{sec:#1}}
\newcommand{\reffig}[1]{Figure~\ref{fig:#1}}
\newcommand{\reflst}[1]{Listing~\ref{lst:#1}}
\newcommand{\reftab}[1]{Table~\ref{tab:#1}}
\newcommand{\refline}[1]{Line~\ref{lst:#1}}
\newcommand{\reflines}[2]{Line~\ref{lst:#1}-\ref{lst:#2}}
\newcommand{\refliness}[2]{Lines~\ref{lst:#1}-\ref{lst:#2}}

\definecolor{bgcolor}{rgb}{0.90,0.90,0.90}
\definecolor{basiccolor}{rgb}{0,0,0}
\definecolor{keywordcolor}{rgb}{0.0,0.0,0.0}
\definecolor{commentcolor}{rgb}{0,0.3,0}
\definecolor{stringcolor}{rgb}{0.35,0.35,0.8}
\definecolor{hicolor}{rgb}{0.1,0.5,0.5}
\definecolor{identcolor}{rgb}{0.3,0.3,0.1}

\lstdefinelanguage{WAWK}{
  morekeywords={when, in-group, in-groups, groups, print, reval, resolve-group, step, load, import, alias, call, if,map,mapa, function},
  otherkeywords={&&,!=,==,<,>,>=,<=,||,@,\#,BEGIN,END,+,=},
  keywordstyle=\color{blue}\bfseries,
  keywordstyle = [2]{\color{red}},
  keywordstyle = [3]{\color{orange}},
  keywordstyle = [4]{\color{green}},
  keywordstyle = [5]{\color{orange}},  
  morekeywords = [2]{@},
  morekeywords = [3]{\#},
  morekeywords = [5]{BEGIN, END},
  keepspaces=true,
  identifierstyle=\color{black},
  comment=[l]{//},
  commentstyle=\color{purple}\ttfamily\it,
  alsoletter={-,;},
  stringstyle=\color{stringcolor}\ttfamily,
  morestring=[b]"
}

\lstset{
  xleftmargin=0.5cm,
  tabsize=2,
  language=WAWK,
  escapeinside={(*@}{@*)},
  basicstyle=\color{basiccolor}\scriptsize\ttfamily,
  breaklines=true,
  postbreak=\mbox{\textcolor{red}{$\hookrightarrow$}\space},
  frame=single,
  firstnumber=1,
  numbers=left,
}

\title{Programming Language Assisted Waveform Analysis:\\A Case Study on the Instruction Performance of SERV}

\author{
	\IEEEauthorblockN{Lucas Klemmer \hspace{1cm} Daniel Gro{\ss}e}
	\IEEEauthorblockA{
		Institute for Complex Systems, Johannes Kepler University Linz, Austria \\
		\{lucas.klemmer, daniel.grosse\}@jku.at
	}\\
        \url{https://wal-lang.org} \hspace{0.5cm} \url{https://github.com/ics-jku/wal/tree/main/wawk}
        \vspace{-7mm}
}

\maketitle

\begin{abstract}
  RISC-V’s growing traction leads to the release of new \mbox{\emph{RISC-V}} cores on a near monthly basis.
  In this growing and diverse ecosystem, understanding the performance and other properties of a RISC-V core is of great importance since selecting the best fitting core
  is mandatory for a successful project.
  Analyzing RISC-V cores by hand is not possible due to the ever-increasing number of available cores and available software benchmarks might not be fine-grained enough to understand a core completely.
  Programming and powerful programming languages have proven to provide the productivity that is required to keep pace with these fast developments.
  
  In this paper we present a case study\footnote{The code for the case study is available at \url{https://github.com/LucasKl/serv-cpi}} in which we use WAWK, a front-end for the open-source \emph{Waveform Analysis Language}, to analyze the performance of all instructions of SERV, a well known bit-serial RISC-V core.
  With WAWK, only a few lines of code are necessary to calculate the respective metric on the waveform generated during simulation.
\end{abstract}

%%%%%%%%%%%%%%%%%%%%%%%%%%%%%%%%%%%%%%%%%%%%%%%%%%%%%%%%%%%%%%%%%%%%%%%%%%%%%%%%%%%%%%%%%%
\section{Introduction}
\label{sec:intro}
RISC-V is an open and royalty free ISA~\cite{riscv-isa1} striving for innovation through collaboration, thus enabling even small companies as well as community projects to develop their own processors which take advantage from RISC-V’s permissive license and its extensibility to explore new ideas and markets with often highly specialized hardware.
However, this openness and extensibility of RISC-V brings its own set of challenges, since the sheer number of available RISC-V cores, which are often highly configurable and extensible, makes it very hard and time-consuming for both, designers and users, to compare different cores against each other~\cite{10.1145/3457388.3458657}~\cite{bestproc}.

In this paper, we use \emph{Waveform AWK} (WAWK), a front-end for
the open-source \emph{Waveform Analysis Language} (WAL)~\cite{KG:2022,KG:2022c}.
WAWK programs have direct access to all signal values of a waveform.
Accessing signals in a WAWK program is similar to accessing variables in regular programming languages with the difference that the value returned depends on the loaded waveform and the time at which the signal is accessed.
Since WAWK is compiled down to WAL it has access to a very rich feature set that provides a large collection of functions which can be used to analyze waveforms.
Thus WAWK allows creating analysis programs using the values from the VCD waveforms generated during simulation of a RISC-V core.

With WAWK, we show that the analysis of RISC-V cores is possible with only a handful of lines which give developers crucial information about the performance of their or others cores.

%%%%%%%%%%%%%%%%%%%%%%%%%%%%%%%%%%%%%%%%%%%%%%%%%%%%%%%%%%%%%%%%%%%%%%%%%%%%%%%%%%%%%%%%%%
\section{Related Work}
\label{sec:related}
In the context of processor architecture research several processor simulators have been proposed.
A prominent example is gem5~\cite{Binkert:2011} or multi2sim~\cite{4384043}.
A complimentary direction are emulators, such as qemu~\cite{qemu} or OVPSim~\cite{ovpsim}.
Both simulators, and emulators can partially be used to calculate (performance) metrics.
However, they are not as flexible as a programmable approach such as WAWK for the problems considered.
WAWK allows developers to create programs and tailor them to the exact requirements, such as IPC count, pipeline analysis and more~\cite{KJG:2022}.
\vspace{-3mm} 
%%%%%%%%%%%%%%%%%%%%%%%%%%%%%%%%%%%%%%%%%%%%%%%%%%%%%%%%%%%%%%%%%%%%%%%%%%%%%%%%%%%%%%%%%%
\section{Preliminaries}
\label{sec:prelim}

%%%%%%%%%%%%%%%%%%%%%%%%%%%%%%%%%%%%%%%%%%%%%%%%%%%%%%%%%%%%%%%%%%%%%%%%%%%%%%%%%%%%%%%%%%
\subsection{WAWK}
\label{sec:framework}
In this paper we use \mbox{WAWK}, a language inspired from the popular text-processing language \emph{AWK}, that is compiled down to WAL.
In general, all WAWK scripts consist of multiple statements that follow a \mbox{\tt condition: \{ action \}} scheme.
For each time index in a waveform, WAWK evaluates the {\tt condition} of each statement, and, if satisfied, executes the associated {\tt action}.

\subsection{SERV}
\label{sec:serv}
SERV is a bit-serial implementation of the RISC-V ISA with a heavy focus on very low area usage.
The bit-serial implementation dictates that all instructions are executed bit by bit such that for example the RV32I \textit{add} instruction is split over at least 32 cycles, one for each bit of the result.

\begin{figure}[tb]
  \centering
  \vspace{5mm}
  \includegraphics[width=0.49\textwidth]{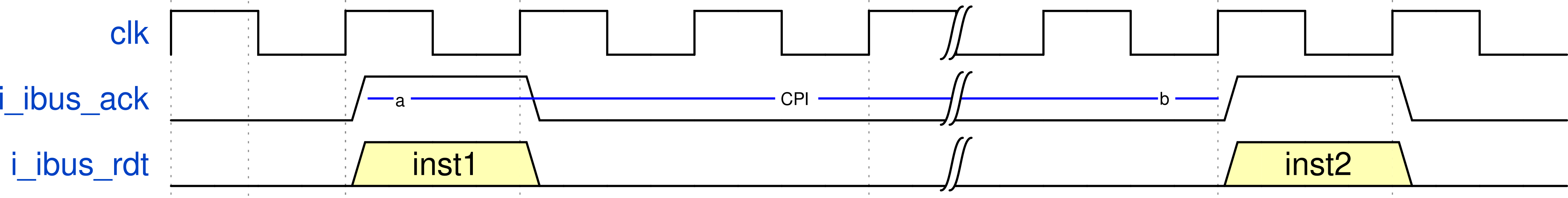}
  \caption{SERV Instuction Lifecycle}
  \label{fig:serv}
  \vspace{-5mm}
\end{figure}

To analyze the runtime of individual instructions in the SERV core we can observe the values of the instruction fetch bus.
The start of a new instruction is signalized by the core when the signal {\tt i\_ibus\_ack} rises.
Therefore, the runtime of one instruction can be calculated by observing two following rising edge events on {\tt i\_ibus\_ack} and calculating the time difference $event_2 - event_1$ between those two events.
This is shown in~\reffig{serv} where the first instruction starts at time {\tt a}.
At time point {\tt b} the next instruction starts thus the {\tt CPI} for this instruction can be calculated by the difference between {\tt b} and {\tt a}.

%%%%%%%%%%%%%%%%%%%%%%%%%%%%%%%%%%%%%%%%%%%%%%%%%%%%%%%%%%%%%%%%%%%%%%%%%%%%%%%%%%%%%%%%%%
\section{CPI Analysis}
\label{sec:experiments}	

The WAWK program for the CPI analysis is shown in~\reflst{code}.
The analysis program consists of four WAWK statements.
First, after the program is started and before the waveform analysis is started the first statement (\refline{code-begin}) is executed once.
This is triggered by the \lstinline{BEGIN} condition which is a special variable which is only true after the program starts.
This WAWK program uses the \emph{riscv-model} Python package to decode RISC-V instructions.
Therefore, an external Python script containing a few lines of Python glue code is imported (\refline{code-import}) and the list of results which will store all measured CPI values is initialized.
Additionally, some aliases are created for often used signals.
Next, some aliases are declared which can be used instead of the long full signal names.

The second statement (\refline{code-count}) is executed at the last cycle of each instruction.
However, the statement is only executed when the opcode of the current instruction matches the opcode which we want to analyze.
This opcode has to be passed to the program as an command line argument and it is available to the WAWK program in the {\tt args} variable.
At this point, the difference to the starting time of this instruction is calculated and added to the list of measured CPI values.

The next statement (\refline{code-new}) is executed whenever a new instruction starts executing.
At this moment, the current time index, which is used to calculate the runtime of this instruction in the previous statement, is stored and the current opcode is decoded and stored in the {\tt op} variable.

Finally, after all indices were processed by the program, the final results, which consist of the average, minimum, and maximum runtimes, are printed by the last statement (\refline{code-end}).

\begin{lstlisting}[caption=WAWK code for CPI the analysis,captionpos=b,label={lst:code}]
BEGIN: { (*@ \label{lst:code-begin} @*)
  import(extern); (*@ \label{lst:code-import} @*)
  cpis = [];
  alias(clk, TOP.servant_sim.dut.cpu.clk);
  alias(fire, TOP.servant_sim.dut.cpu.i_ibus_ack);
  alias(instruction, TOP.servant_sim.dut.cpu.i_ibus_rdt);
}

clk, !fire, fire@2, op == args[0]: { (*@ \label{lst:code-count} @*)
  cpis = cpis + ((INDEX - start) / 2);
}

clk, fire: { (*@ \label{lst:code-new} @*)
  start = INDEX;
  op = call(extern.decode, instruction);
}

END: { (*@ \label{lst:code-end} @*)
  if (cpis) {
    if (min(cpis) == max(cpis)) {
      printf("%10s %10d\n", args[0], average(cpis));
    } else {
      printf("%10s %10d %10d %10d\n", args[0], average(cpis), min(cpis), max(cpis));
    };
  };
}
\end{lstlisting}

\section{Results}
\label{sec:results}

\begin{table}[tb]
  \centering
  \begin{tabular}{lrrr}
    \toprule
    Opcode &       Avg. &        Min &        Max \\
    \midrule
     auipc &         35 &            &            \\
       lui &         35 &            &            \\
       add(i) &         35 &            &            \\
       sub &         35 &            &            \\
       and(i) &         35 &            &            \\
        or(i) &         35 &            &            \\
       xor(i) &         35 &            &            \\
     ecall &         35 &            &            \\
       slt(i) &         68 &            &            \\
      sltu &         68 &            &            \\
   sltiu &         68 &            &            \\
       sll(i) &         68 &            &            \\
       sra &         75 &         68 &         99 \\
      srai &         70 &         68 &         99 \\
       srl(i) &         75 &         68 &         99 \\
       jal &         68 &         68 &         70 \\
      jalr &         69 &         68 &         70 \\
       beq &         68 &         68 &         70 \\
       bge &         69 &         68 &         70 \\
      bgeu &         69 &         68 &         70 \\
       blt &         68 &         68 &         70 \\
      bltu &         69 &         68 &         70 \\
       bne &         68 &         68 &         70 \\
        lb &         69 &            &            \\
        lh &         69 &         69 &         70 \\
       lhu &         69 &         69 &         70 \\
        lw &         69 &         69 &         70 \\
        sh &         69 &         69 &         70 \\
        sw &         69 &         69 &         70 \\
    \bottomrule
  \end{tabular}
  \caption{SERV CPI on Compliance Tests}
  \label{tab:cpis}
  \vspace{-5mm}
\end{table}

\reftab{cpis} shows the analysis results for each of the instructions of the SERV core.
The first column lists the name of the instruction, the second lists the average number of cycles this instruction requires to be executed, and the last two columns show the minimum and maximum number of instructions required to execute this instruction respectively.
The table shows that the instructions mainly fall into two categories.
One half of the instructions always finishes execution in constant time while the over half requires a variable number of cycles.

%%%%%%%%%%%%%%%%%%%%%%%%%%%%%%%%%%%%%%%%%
\section{Conclusion}
\label{sec:conclusions}

In this paper we presented a programmable analysis of the performance of individual instructions as implemented in the well-known RISC-V core SERV.
The analysis program is written in WAWK, a DSL for waveform analysis inspired by the AWK text-processing language.
In only a few lines of code the program analyzes the number of cycles per instruction required to execute a specific instruction.

\vspace{-3mm}
\section*{Acknowledgments}
This work has partially been supported by the LIT Secure and Correct Systems Lab funded by the State of Upper Austria.
\vspace{-2mm}

\bibliographystyle{IEEEtran}

\bibliography{lit,lit_header_long}

\end{document}